%% file: inspiral_svd.tex
\documentclass[aps,prd,twocolumn,showpacs,unsortedaddress,superscriptaddress,showkeys,preprintnumbers,letterpaper]{revtex4-1}
\pdfoutput=1
\input{packages}

\newcommand{\mat}[1]{{\bf #1}}
\newcommand*{\ee}{\mathrm{e}}
\newcommand*{\aye}{\mathrm{i}}
\newcommand*{\diff}{\,\mathrm{d}}
\newcommand{\norm}[1]{\left\lVert #1 \right\rVert}
\newcommand{\abs}[1]{\left\lvert #1 \right\rvert}

\begin{document}

\newcommand{\LIGOCaltech}{LIGO Laboratory, California Institute of Technology, Pasadena, CA 91125, USA}
\newcommand{\TAPIR}{Theoretical Astrophysics, California Institute of Technology, Pasadena, CA 91125, USA}

\title{
Singular value decomposition applied to compact binary coalescence gravitational-wave signals
}

\date{\today}

\author{Kipp~Cannon} 
\email{kipp.cannon@ligo.org}
\affiliation{\LIGOCaltech}

\author{Adrian~Chapman} 
\affiliation{\LIGOCaltech}

\author{Chad~Hanna}  
\email{chad.hanna@ligo.org}
\affiliation{\LIGOCaltech}

\author{Drew~Keppel}  
\email{drew.keppel@ligo.org}
\affiliation{\LIGOCaltech}
\affiliation{\TAPIR}

\author{Antony~C.~Searle}
\email{antony.searle@ligo.org}
\affiliation{\LIGOCaltech}

\author{Alan~J.~Weinstein}
\email{ajw@caltech.edu}
\affiliation{\LIGOCaltech}

\begin{abstract}

We investigate the application of the singular value decomposition to
compact-binary, gravitational-wave data-analysis. We find that the truncated
singular value decomposition reduces the number of filters required to analyze
a given region of parameter space of compact binary coalescence waveforms by an
order of magnitude with high reconstruction accuracy. We also compute an
analytic expression for the expected signal-loss due to the singular value
decomposition truncation.

\end{abstract}



\preprint{}

\maketitle

\input{acros}

\section{Introduction}
\label{sec:introduction}

The coalescence of compact binaries composed of neutron stars and or black
holes is a promising source of gravitational radiation for ground-based \ac{GW}
detectors.  The mass parameters of the \ac{GW} signal are not known \emph{a
priori}. In order to detect \ac{GW} from \ac{CBC} events, a large number of
filter templates are required to to probe the continuous component mass
parameter space, $(m_1,m_2)$, of possible \ac{CBC} signals in the detector data
to high fidelity~\cite{Owen:1995tm, Owen:1998dk}.  Template waveforms are
distributed in the space such that there is a small maximum loss of \ac{SNR}
(called the ``minimal match") due to the mismatch between an arbitrary point in
the mass parameter space and the nearest discrete point of the template bank.
A standard choice for the minimal match is 97\%, which, for a hexagonally
tiled, flat, two-dimensional manifold, corresponds to neighboring templates
that have greater than 95\% overlap.

This redundancy implies that correlated calculations are required to filter the
data with these templates.  The \ac{SVD} can be used to eliminate these
correlations by producing orthogonal basis vectors that can be used for
filtering and reconstructing the original template bank.

This work will describe how to reduce the computational redundancy in filtering
the \ac{CBC} signal parameter-space in order to more efficiently infer whether
or not a \ac{GW} is present.  Specifically, we will explore a purely numerical
technique using the \ac{SVD} to reduce the number of templates required to
search the data.  We note that others have applied the use of \ac{SVD} to
\ac{GW} data-analysis to analyze optimal \ac{GW} burst detection
\cite{bradyraymajumder2004, heng2008} and coherent networks of detectors
\cite{wen2008}.  We also note that significant work has been done to
analytically reduce the computational filtering burden using interpolation for
certain template waveforms~\cite{Croce2000,finn2005}.

This paper is organized as follows. First, we describe the framework for
\ac{CBC} filtering in the context of vector inner products.  Next we introduce
the \ac{SVD} as a way to reduce the number of filters required to approximately
compute those inner products.  We then derive an expression for the expected
\ac{SNR} loss in terms of the singular values.  Finally, we demonstrate the
application of this method to a set of \ac{CBC} waveforms corresponding to
\ac{BNS} coalescences.

\section{Method}
\label{sec:method}

\subsection{Matched filtering}

\ac{CBC} searches employ matched filtering as the first step in locating a
\ac{GW} signal~\cite{findchirppaper}.  The optimal filtering strategy weights
both the detector output and template waveform by the inverse of the amplitude
spectral density of the detector noise, a process called ``whitening''.
Representing both the whitened data and the $\alpha^{\mathrm{th}}$ whitened
template waveform as discretely sampled time series, $\vec{s} = \{s_{i}\}$ and
$\vec{h}_{\alpha} = \{{h_{\alpha}}_{i}\}$, respectively, the output of the
matched filter at a specific point in time is given by the vector inner product 
\begin{equation}
\rho_{\alpha} = \vec{h}^*_{\alpha} \cdot \vec{s} \,.  
\end{equation} 
In searches for \ac{GW}s from \ac{CBC} sources, the signals being sought are
chirping sinusoids with an unknown phase.  The search over phase is
accomplished through the use of complex-valued templates where
$\Re\vec{h}_{\alpha}$ contains the cosine-like phase and $\Im\vec{h}_{\alpha}$
contains the sine-like phase.  The filter output can be maximized over template
phase by evaluating $\abs{\rho_{\alpha}}$.

In the absence of a \ac{GW} signal, the whitened detector data consists only of
noise, $\vec{n} = \{n_{i}\}$, and is a stationary, zero-mean, unit-variance,
Gaussian random process, so
\begin{subequations}
\label{eqn1}
\begin{gather}
\langle n_i \rangle = 0 \,, \\
\langle n_i n_j \rangle = \delta_{ij} \,,
\end{gather}
\end{subequations}
where $\langle\,\,\rangle$ denotes the ensemble average.  When the template
waveforms are normalized such that $\Re\vec{h}_{\alpha} \cdot
\Re\vec{h}_{\alpha} = \Im\vec{h}_{\alpha} \cdot \Im\vec{h}_{\alpha} = 1$,
\eqref{eqn1} yields
\begin{subequations}
\label{eqn2}
\begin{gather}
\langle \vec{h}^*_{\alpha} \cdot \vec{n} \rangle = 0 \,, \\
\langle ( \Re\vec{h}_{\alpha} \cdot \vec{n} )^2 \rangle = \langle (
\Im\vec{h}_{\alpha} \cdot \vec{n} )^2 \rangle = 1 \,.
\end{gather}
\end{subequations}
When \eqref{eqn2} is true, $\rho_{\alpha}$ is called the \ac{SNR} and indicates
how likely it is that a signal is present in the data at that point in
time~\cite{wainstein:1962}.

As explained in Sec.~\ref{sec:introduction}, $\vec{h}_{\alpha} \cdot
\vec{h}_{\alpha'} > 0.95$ for adjacent templates.  For those templates,
$\rho_{\alpha}$ and $\rho_{\alpha'}$ differ by at most 5\%. This suggests the
existence of an approximation scheme that would allow the \ac{SNR}s to be
computed to reasonable accuracy without explicitly evaluating all the template
inner products.  Next, we will look at how the truncated \ac{SVD} can be used
to replace the template bank with an approximate, lower-rank, orthogonal basis
from which the \ac{SNR}s can be reconstructed.

\subsection{Reducing the number of filters with truncated singular value decomposition}

\begin{figure}
\includegraphics{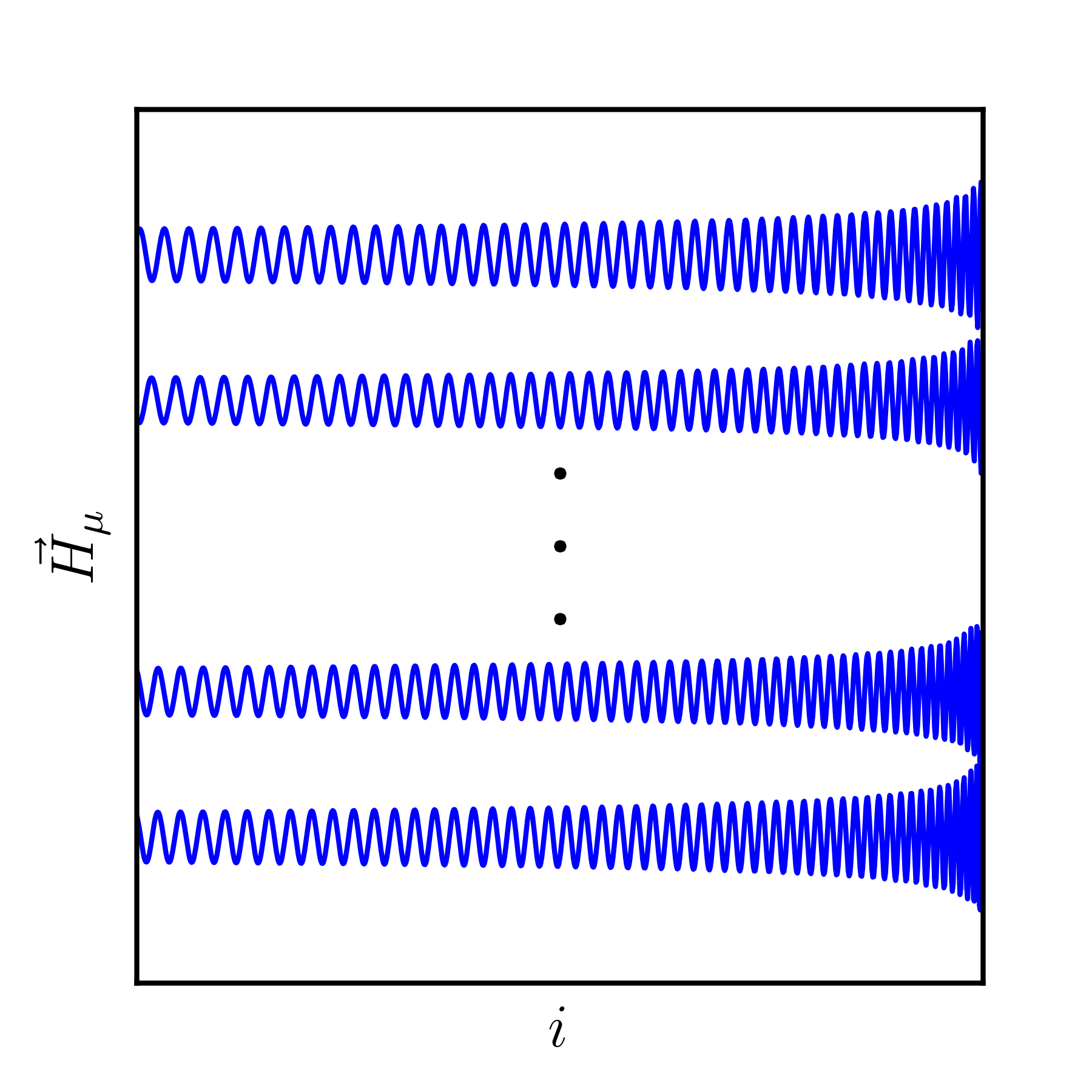}
\includegraphics{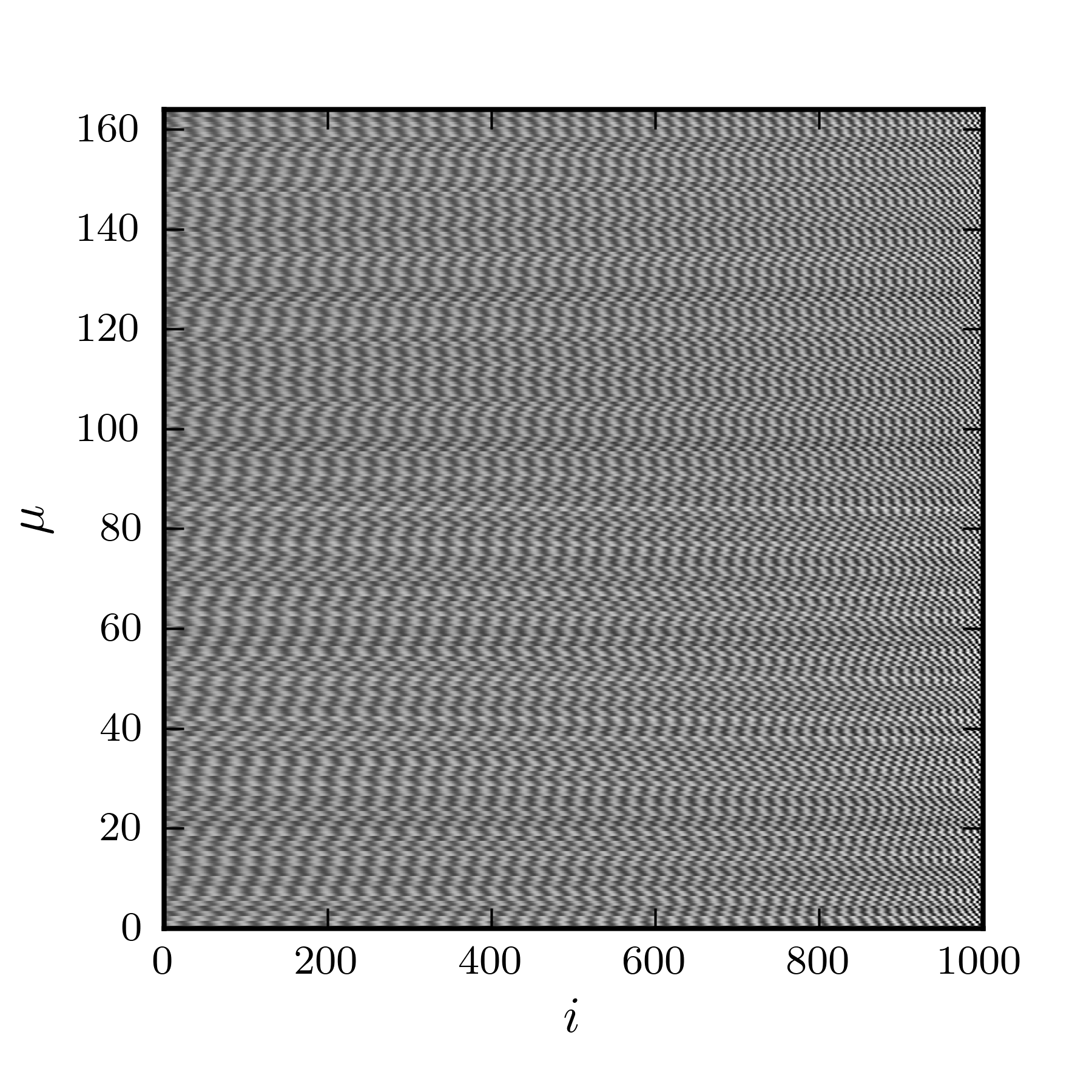}
\caption{An example template matrix, $\mat{H}$. Top: An illustration of how the
input template time series is packed into the template matrix. Bottom: The
matrix of the template time series where the y-axis indicates the template
waveform and the x-axis indicates the time samples.  It should be noted that
these waveforms have been shortened and have not been whitened for illustrative
purposes.}
\label{fig:templatematrix}
\end{figure}

The waveforms are parameterized by their component masses and we denote the
$\alpha^{\rm th}$ template waveform of the $M$ templates required to search a
given parameter space as $\vec{h}_{\alpha} = \{ h(m_{1}, m_{2}, t_{i}) \}$.
Rather than filter the data with $N=2M$ real-valued filters ($M$ complex-valued
filters), we linearly combine the output of a basis set of fewer, real-valued,
filters, $\vec{u}_\mu$, to reproduce $\rho_{\alpha}$ to the desired accuracy,
$\rho'_{\alpha}$.  The goal is to have
\begin{equation}
\label{eq:linear_combination_of_snr}
\rho'_{\alpha} = \sum_{\mu=1}^{N'} A_{\alpha \mu} (\vec{u}_\mu \cdot
\vec{s}) \,,
\end{equation}
where $\mat{A}$ is the complex-valued reconstruction matrix we wish to find and
the number of inner products is reduced from $N$ to $N'$. In order to find the
basis vectors, $\vec{u}_\mu$, we use the \ac{SVD} of the real-valued template
matrix, $\mat{H}$
\begin{multline}
\label{eqn:Hpacking}
\mat{H} = \{H_{\mu j}\} \\
= \{ \Re\vec{h}_{1}, \Im\vec{h}_{1}, \Re\vec{h}_{2},
\Im\vec{h}_{2}, ..., \Re\vec{h}_{M}, \Im\vec{h}_{M}\} \,,
\end{multline}
where $\mu$ identifies rows of $\mat{H}$ and indexes the filter number, and $j$
identifies the columns of $\mat{H}$ and indexes sample points. In this
definition, the row vectors $\vec{H}_{(2\alpha-1)}$ and $\vec{H}_{(2\alpha)}$
are, respectively, the real and imaginary parts of the $\alpha^{\rm th}$
complex waveform, $\vec{h}_{\alpha}$. An illustrative template matrix can be
seen in Fig.~\ref{fig:templatematrix}.

\begin{figure}
\includegraphics{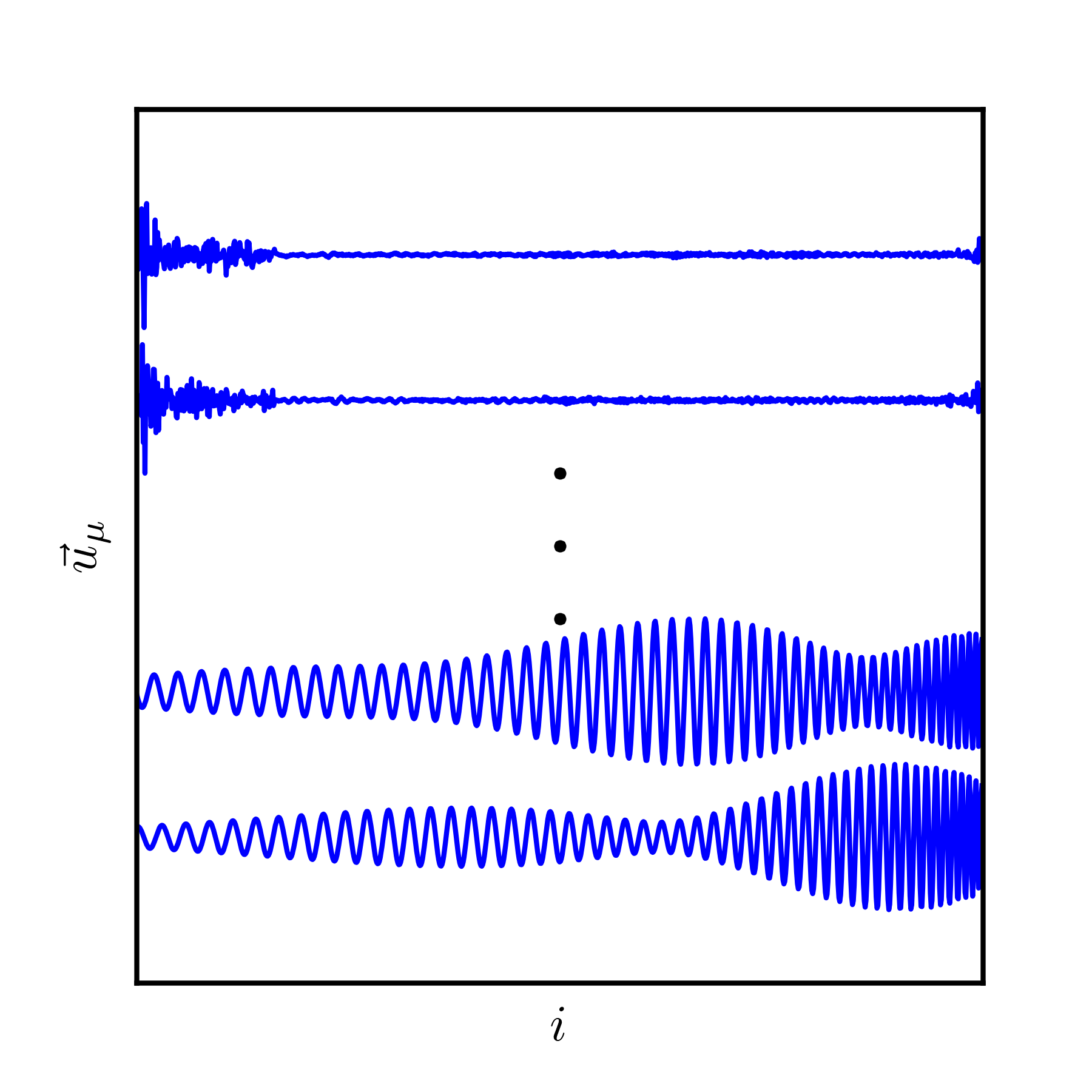}
\includegraphics{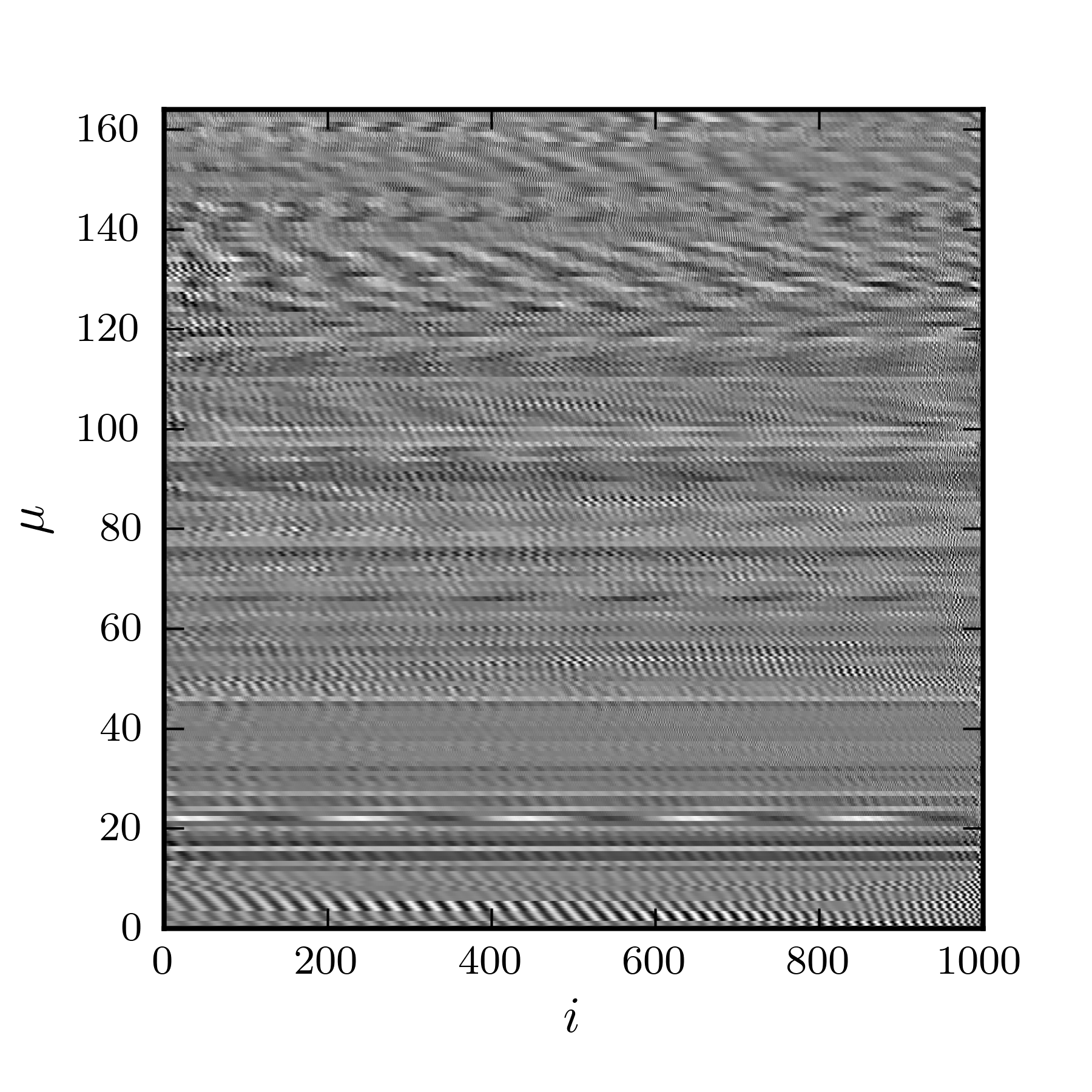}
\caption{An example basis matrix, $\mat{u}$. Top: An illustration of the
resulting orthonormal basis vectors ordered from most to least important
(bottom to top) in reconstructing $\mat{H}$.  Bottom: The matrix of basis
waveforms produced by the \ac{SVD}.  The y-axis indexes the basis waveforms and
the x-axis indicates time samples. It should be noted that these basis vectors
have been computed from shortened, non-whitened template waveforms as mentioned
in Fig.~\ref{fig:templatematrix} purely for illustrative purposes.}
\label{fig:basismatrix}
\end{figure}

The \ac{SVD} factors a matrix such that~\cite[Sec. 14.4]{GSL}
\begin{equation}
\label{eqn:svd}
H_{\mu j} = \sum\limits_{\nu=1}^N v_{\mu \nu} \sigma_{\nu} u_{\nu j} \,,
\end{equation}
where $\mat{v}$ is an orthonormal matrix of reconstruction coefficients
whose columns, $v_{\mu\nu}$, satisfy
\begin{equation}
\label{eqn:vs}
\sum_\mu v_{\mu\nu} v_{\mu\lambda} = \delta_{\nu\lambda} \,,
\end{equation}
$\vec{\sigma}$ is a vector of singular values ranked in order of importance in
reconstructing the $\mat{H}$, and $\mat{u}$ is a matrix of orthonormal bases
(e.g.\ an illustration can be found in Fig.~\ref{fig:basismatrix}) whose rows
are basis vectors, $\vec{u}_{\mu}$, satisfying
\begin{equation}
\label{eqn:us}
\sum_j u_{\mu j} u_{\nu j} = \delta_{\mu\nu} \,.
\end{equation}

However, since a search for \ac{CBC} signals only needs waveform accuracies of
a few percent to be successful, it is possible to make an approximate
reconstruction of $\mat{H}$
\begin{equation}
\label{eq:svdapprox}
H_{\mu j} \approx H'_{\mu j} :=
\sum_{\nu=1}^{N'} v_{\mu \nu} \sigma_{\nu} u_{\nu j} \,,
\end{equation}
where $N' < N$. This reduces the number of rows of $\mat{u}$ used in the
reconstruction. We create a new basis matrix $\mat{u} = \{u_{\nu j}\} =
\{\vec{u}_1, \vec{u}_2, ...\vec{u}_{N'}\}$, where $\nu$ indexes the filter
number, $j$ indexes sample points, and we have discarded the basis vectors that
look least like the template waveforms (i.e.\ with the lowest singular values).
We can write \eqref{eq:linear_combination_of_snr} as
\begin{align}
\label{eq:svdsnr}
\rho'_{\alpha} &= \left( \vec{H}'_{(2\alpha-1)} - \aye \vec{H}'_{(2\alpha)} \right) \cdot \vec{s} \nonumber\\
&=\sum_{\nu=1}^{N'} \left( v_{(2\alpha - 1) \nu} \sigma_{\nu} - \aye v_{(2\alpha) \nu} \sigma_{\nu} \right)
\left(\vec{u}_\nu \cdot \vec{s} \right) \,,
\end{align}
where we have made use of the packing of $\mat{H}$ \eqref{eqn:Hpacking} and
\eqref{eq:svdapprox}.

\subsection{Reconstruction accuracy}

As we are not reconstructing the original template waveforms exactly, there will
be some inherent mismatch between $\vec{H}'_\mu$ and $\vec{H}_\mu$. We want to
know the expected fractional \ac{SNR} we will lose because of this difference.

As stated previously, the inner product of a (normalized) template waveform,
$\vec{H}_\mu$, with itself is
\begin{equation}
\label{eqn:normalized}
\vec{H}_\mu \cdot \vec{H}_\mu = 1 = \sum_{\nu=1}^{N} v_{\mu\nu}^2 \sigma_\nu^2
\,,
\end{equation}
where, in the second line, we have made use of the orthogonality of basis
vectors \eqref{eqn:us}.  A similar relation can be found for the inner product
of the reconstructed waveform, $\vec{H}'_\mu$, with itself
\begin{equation}
\label{eqn:discarded}
\vec{H}'_\mu \cdot \vec{H}'_\mu = \sum_{\nu=1}^{N'} v_{\mu\nu}^2 \sigma_\nu^2 =
1 - \sum_{\nu=N'+1}^{N} v_{\mu\nu}^2 \sigma_\nu^2  \,.
\end{equation}
Because of the orthogonality of the basis vectors \eqref{eqn:us}, the inner
product between a template waveform, $\vec{H}_\mu$, with a reconstructed
waveform, $\vec{H}'_\nu$, is
\begin{equation}
\label{eqn:innersimple}
\vec{H}_\mu \cdot \vec{H}'_\nu = \vec{H}'_\mu \cdot \vec{H}_\nu =
 \vec{H}'_\mu \cdot \vec{H}'_\nu \,.
\end{equation}

In addition, the two phases of the templates, which are packed adjacently in
$\mat{H}$ \eqref{eqn:Hpacking}, are orthogonal
\begin{equation}
\label{eqn:H0.H1}
\vec{H}_{(2\mu-1)} \cdot \vec{H}_{(2\mu)} =
 \sum_{\nu=1}^{N} v_{(2\mu-1)\nu}v_{(2\mu)\nu} \sigma_{\nu}^2 =  0 \,.
\end{equation}
This implies that the inner product of the two phases of the approximate
waveforms are given as
\begin{align}
\label{eqn:H'0.H'1}
\vec{H}'_{(2\mu-1)} \cdot \vec{H}'_{(2\mu)}
&= \sum_{\nu=1}^{N'} v_{(2\mu-1)\nu}v_{(2\mu)\nu} \sigma_{\nu}^2 \nonumber\\
&= -\sum_{\nu=N'+1}^{N} v_{(2\mu-1)\nu}v_{(2\mu)\nu} \sigma_{\nu}^2 \,.
\end{align}

The average fractional \ac{SNR} loss, $\delta\rho_\alpha / \rho_\alpha$,
between a template waveform and the two phases of the same reconstructed
waveform is given by
\begin{equation}
\label{eqn:fracsnrloss}
\frac{\delta\rho_\alpha}{\rho_\alpha} :=
 1 - \frac{\abs{\rho'_\alpha}}{\abs{\rho_\alpha}} \,.
\end{equation}

The following derives the mismatch in terms the of components we truncate from
the \ac{SVD}. First we compute these terms for a given signal waveform, $\vec{s}
= \Re \left(A\ee^{\aye \phi} \vec{h}_{\alpha}\right)$, with phase, $\phi$.  The
\ac{SNR} from the exact waveform, $\abs{\rho_\alpha(\phi)}$, is given as
\begin{multline}
\abs{\rho_\alpha(\phi)} =
\left[ \left( \frac{\Re\vec{h}_{\alpha} \cdot
A\Re \left(\ee^{\aye \phi} \vec{h}_{\alpha}\right)}
{\sqrt{\Re\vec{h}_{\alpha} \cdot \Re\vec{h}_{\alpha}})} \right)^2 \right.\\
\left.+\left( \frac{\Im\vec{h}_{\alpha} \cdot
A\Re \left(\ee^{\aye \phi} \vec{h}_{\alpha}\right)}
{\sqrt{\Im\vec{h}_{\alpha} \cdot \Im\vec{h}_{\alpha}})} \right)^2 \right]^{1/2}
= A \,,
\end{multline}
in which we have used \eqref{eqn:normalized} and \eqref{eqn:H0.H1}. The
\ac{SNR} from the approximate waveform, $\abs{\rho'_\alpha}$, is given as
\begin{multline}
\label{eqn:rho'}
\abs{\rho'_\alpha(\phi)} =
\left[ \left( \frac{\Re\vec{h}'_{\alpha} \cdot
A\Re \left(\ee^{\aye \phi} \vec{h}_{\alpha}\right)}
{\sqrt{\Re\vec{h}'_{\alpha} \cdot \Re\vec{h}'_{\alpha}})} \right)^2 \right.\\
\left.+\left( \frac{\Im\vec{h}'_{\alpha} \cdot
A\Re \left(\ee^{\aye \phi} \vec{h}_{\alpha}\right)}
{\sqrt{\Im\vec{h}'_{\alpha} \cdot \Im\vec{h}'_{\alpha}})} \right)^2 \right]^{1/2}
\end{multline}
%
We can expand \eqref{eqn:rho'} using the packing of $\mat{H}$
\eqref{eqn:Hpacking}, \eqref{eqn:discarded}, \eqref{eqn:innersimple}, and
\eqref{eqn:H'0.H'1} to 
\begin{multline}
\label{eqn:snr'raw}
\abs{\rho'_\alpha(\phi)} = A \left[ \cos^2\phi \left(1 - \sum_{\mu=N'+1}^{N} v_{(2\alpha-1)\mu}^2 \sigma_{\mu}^2 \right) \right.\\
+ \sin^2\phi \left(1 - \sum_{\mu=N'+1}^{N} v_{(2\alpha)\mu}^2 \sigma_{\mu}^2 \right) \\
+ 4 \cos\phi\sin\phi \sum_{\mu=N'+1}^{N} v_{(2\alpha-1)\mu} v_{(2\alpha)\mu} \sigma_{\mu}^2 \\
+ \sin^2\phi \frac{\left( \sum_{\mu=N'+1}^{N} v_{(2\alpha-1)\mu} v_{(2\alpha)\mu} \sigma_{\mu}^2 \right)^2}{1 - \sum_{\mu=N'+1}^{N} v_{(2\alpha-1)\mu}^2 \sigma_{\mu}^2} \\
\left.+ \cos^2\phi \frac{\left( \sum_{\mu=N'+1}^{N} v_{(2\alpha-1)\mu} v_{(2\alpha)\mu} \sigma_{\mu}^2 \right)^2}{1 - \sum_{\mu=N'+1}^{N} v_{(2\alpha)\mu}^2 \sigma_{\mu}^2} \right]^{1/2} \,.
\end{multline}

Let us look at the higher order sums in \eqref{eqn:snr'raw}. The sums
$\sum_{\nu=N'+1}^{N} v_{\mu\nu}^2 \sigma_{\nu}^2$, which are also found in
\eqref{eqn:discarded}, represent the power of vector $\vec{H}_\mu$ lost through
the truncation of the \ac{SVD}. These sums must be less than 1,
$\sum_{\nu=N'+1}^{N} v_{\mu\nu}^2 \sigma_{\nu}^2 < 1$.  However, since the
objective is for the approximation to be such that $\norm{\vec{H}_\mu -
\vec{H}'_\mu} \sim 1\%$, we expect
\begin{equation}
\sum_{\nu=N'+1}^{N} v_{\mu\nu}^2 \sigma_{\nu}^2 \ll 1 \,,
\end{equation}
and we can therefore drop terms that are higher than first order in these sums.
Additionally,
\begin{multline}
\abs{\sum_{\nu=N'+1}^{N} v_{\mu\nu} v_{\mu'\nu} \sigma_{\nu}^2} \le \\
\sqrt{\left( \sum_{\nu=N'+1}^{N} v_{\mu\nu}^2 \sigma_{\nu}^2 \right)
\left( \sum_{\nu=N'+1}^{N} v_{\mu'\nu}^2 \sigma_{\nu}^2 \right)}
\ll 1 \,.
\end{multline}
This means \eqref{eqn:snr'raw} is approximately
\begin{multline}
\label{eqn:approxsnrloss}
\abs{\rho'_\alpha(\phi)} \approx
A \left[1 - \frac{1}{2}\cos^2\phi \sum_{\mu=N'+1}^{N}
	v_{(2\alpha-1)\mu}^2 \sigma_{\mu}^2  \right.\\
- \frac{1}{2}\sin^2\phi \sum_{\mu=N'+1}^{N}
	v_{(2\alpha)\mu}^2 \sigma_{\mu}^2 \\
\left.+ 2 \cos\phi\sin\phi \sum_{\mu=N'+1}^{N}
	v_{(2\alpha-1)\mu} v_{(2\alpha)\mu} \sigma_{\mu}^2
\right] \,.
\end{multline}

As physical signals will arrive in the detectors with random phases, we now
average over the phase, $\phi$, using
\begin{equation}
\abs{\rho_\alpha} := \frac{1}{2\pi} \int_{0}^{2\pi} \abs{\rho_\alpha(\phi)} \diff\phi \,,
\end{equation}
resulting in
\begin{subequations}
\label{eqn:snravephi}
\begin{gather}
\abs{\rho_\alpha} = A \,, \\
\abs{\rho'_\alpha} =
A \left[1 - \frac{1}{4}\sum_{\mu=N'+1}^{N}
	\left(v_{(2\alpha-1)\mu}^2 + v_{(2\alpha)\mu}^2 \right)\sigma_{\mu}^2
\right] \,.
\end{gather}
\end{subequations}

Substituting \eqref{eqn:snravephi} in \eqref{eqn:fracsnrloss}, we find the
average fractional \ac{SNR} loss for the $\alpha^{\rm th}$ template
\begin{equation}
\label{eqn:fracsnrlosseqn}
\frac{\delta\rho_\alpha}{\rho_\alpha} =
\frac{1}{4}\sum_{\mu=N'+1}^{N}
\left(v_{(2\alpha-1)\mu}^2 + v_{(2\alpha)\mu}^2 \right)\sigma_{\mu}^2 \,.
\end{equation}

The expected fractional \ac{SNR} loss can be computed by averaging over the
waveforms in the template bank using
\begin{equation}
\label{eqn:aveoverbank}
\left\langle \frac{\delta\rho}{\rho} \right\rangle :=
\frac{1}{M} \sum_{\alpha = 1}^{M} \frac{\delta\rho_\alpha}{\rho_\alpha} \,.
\end{equation}
Combining \eqref{eqn:fracsnrlosseqn} with \eqref{eqn:aveoverbank}, remembering
$M = N/2$, and using the orthogonality of reconstruction coefficients
\eqref{eqn:vs}, we get
\begin{equation}
\label{eqn:avefracsnrloss}
\left\langle \frac{\delta\rho}{\rho} \right\rangle  = \frac{1}{2N}\sum_{\mu=N'+1}^{N} \sigma_{\mu}^2 \,.
\end{equation}

It is not surprising that the expected fractional \ac{SNR} loss is proportional
to the square of the Frobenius norm of the truncation error of $\mat{H}$
\begin{equation}
\norm{\mat{H} - \mat{H}'}_{2}^{2} = \sum_{\mu, j} \left( H_{\mu j} - H'_{\mu j}
\right)^2 = \sum_{\nu=N'+1}^{N} \sigma_\nu^2 \,.
\end{equation}

The expected fractional \ac{SNR} loss, $\langle \delta \rho / \rho \rangle$,
can be used as a threshold for deciding how many basis vectors to keep in the
truncated \ac{SVD} reconstruction of the template matrix. For detection
purposes, we want $\langle \delta \rho / \rho \rangle$ to be less than the
minimal match of the template bank.

\section{Application to compact binary coalescence gravitational-wave signals}

We apply the above procedure to \ac{BNS} waveforms with chirp masses $1.125
M_{\odot} \le M_c < 1.240 M_{\odot}$ and component masses $1 M_{\odot} \le
m_1,m_2 < 3 M_{\odot}$. The number of templates required to hexagonally cover
this range in parameters using a minimal match of $96.8\%$ is $M=456$, which
implies a total number of filters $N=912$.  These non-spinning waveforms were
produced to 3.5PN order\cite{LAL}, sampled at 2048 Hz, up to the Nyquist
frequency of 1024 Hz.  The last 10 seconds of each waveform, whitened with the
initial LIGO amplitude spectral density, were used to construct $\mat{H}$.

\begin{figure}
\includegraphics{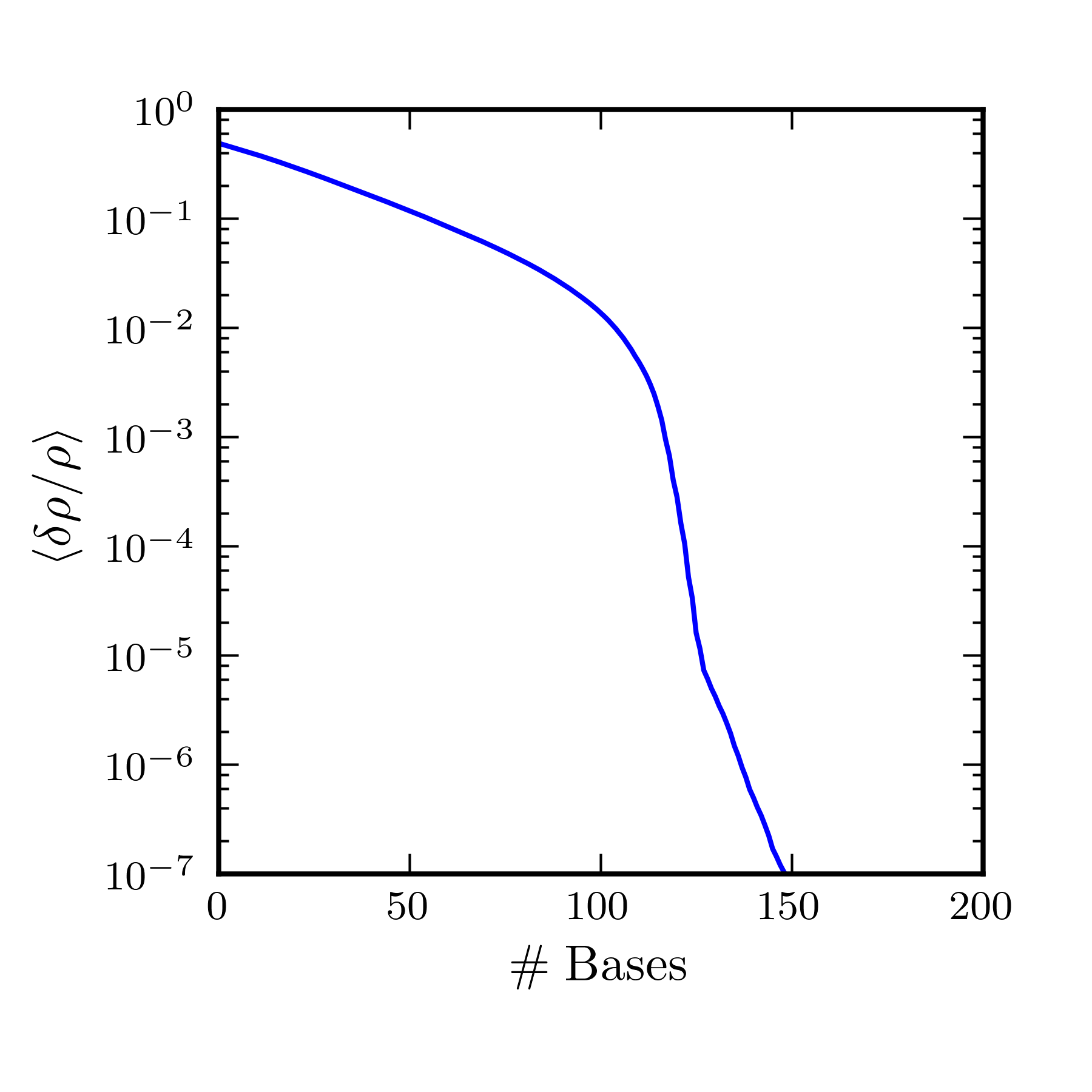}
\caption{The expected fractional \ac{SNR} loss, $\langle \delta \rho / \rho
\rangle$, given by \eqref{eqn:avefracsnrloss} as a function of the number of
basis vectors we retain (out of $N=912$).  The region $\langle \delta \rho /
\rho \rangle > 10\%$ should be ignored as the Taylor expansion of the
fractional \ac{SNR} loss in \eqref{eqn:approxsnrloss} is not valid in that
regime.}
\label{fig:snrloss}
\end{figure}

In Fig.~\ref{fig:snrloss}, we plot $\langle \delta \rho / \rho \rangle$ as a
function of the number of basis vectors kept. If we require that $\langle
\delta \rho / \rho \rangle = 10^{-3}$, we find we can reduce the number of
filters in the above template bank from $N=912$ to $N'=118$, about an order of
magnitude reduction in the number of filters.

\begin{figure}
\includegraphics{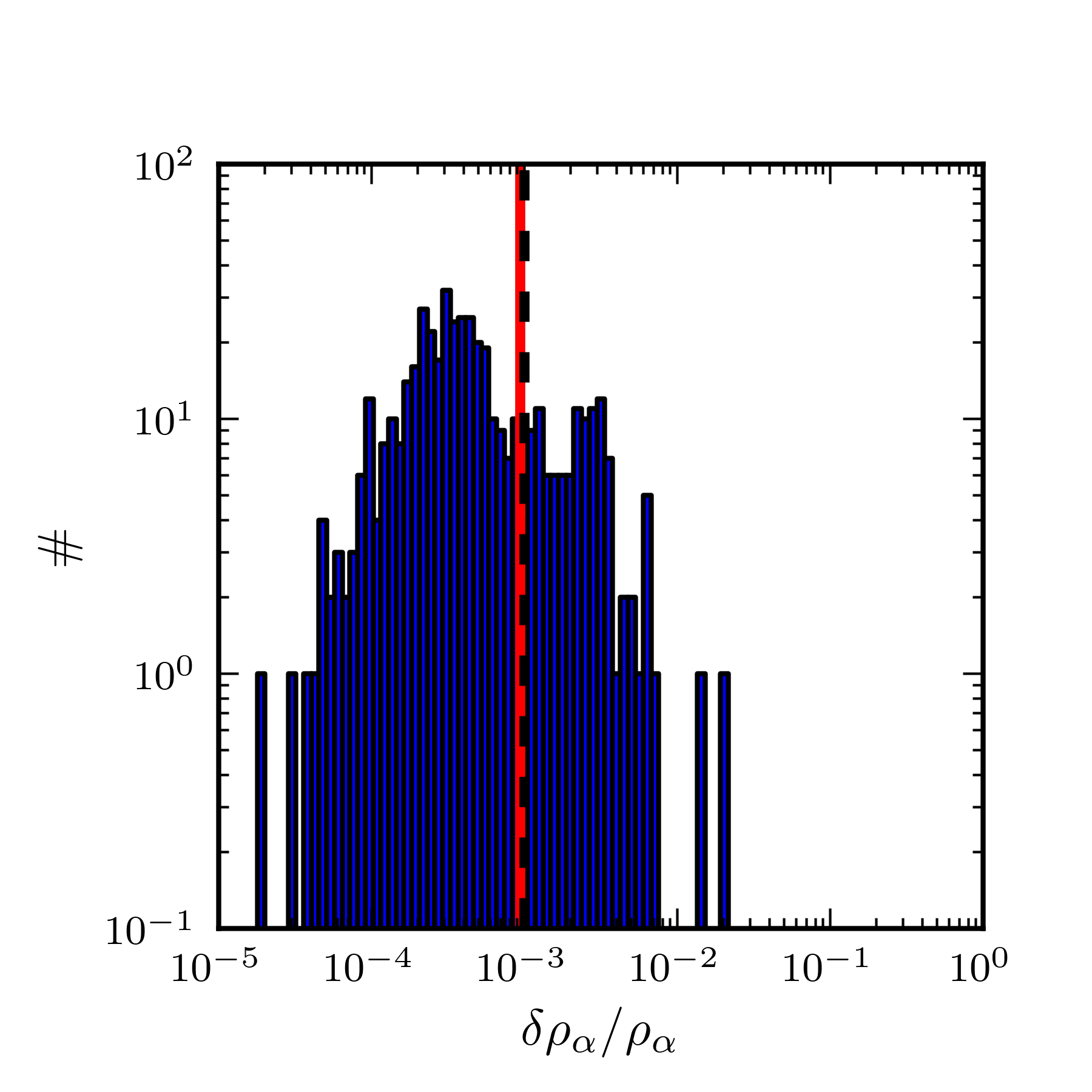}
\caption{Histogram of measured fractional \ac{SNR} loss, $\delta \rho_\alpha /
\rho_\alpha$, where we have chosen a random values of $\phi$ for each template.
The mean value predicted by \eqref{eqn:avefracsnrloss}, shown as the
dashed-black line, matches the measured mean shown in solid-red.}
\label{fig:snrlosshist}
\end{figure}

In Fig.~\ref{fig:snrlosshist} we show how $\langle \delta \rho / \rho \rangle$
compares to the actual distribution of $\delta \rho_\alpha / \rho_\alpha$,
where we have chosen random values of $\phi$ for each template. We find it is a
good measure of the expected fractional loss of \ac{SNR}.

We have investigated how generic this reduction of filters is for other regions
of \ac{CBC} mass parameter space (e.g., regions of parameter space with larger
component masses), and find the reduction to be similar.  We tested this by
generating a template bank with a $96.8\%$ minimal match, component masses
between $1 M_{\odot}$ and $34 M_{\odot}$, and total mass below $35 M_{\odot}$.
We then ordered the templates by chirp mass, split the template bank up into
patches of $M=456$ templates, and computed the \ac{SVD} for these patches.

We can include larger portions of parameter space in the \ac{SVD} by including
more templates such that the number of templates is smaller than the number of
time-samples per template. However the compuational cost of the \ac{SVD} of an
$N \times L$ matrix with $N \le L$ grows as $\mathcal{O}(LN^2)$, thus including
more templates nonlinearly increases the cost. Another complication is that
waveforms further apart in parameter space have smaller overlap. This will
result in more basis vectors being required to reconstruct the waveforms to the
same accuracy. Therefore, including larger portions of parameter space in a
single \ac{SVD} computation will result in diminishing returns for the
computational cost. We propose to address this issue, as above, by breaking up
the parameter space into patches for which we can independently compute the
\ac{SVD}, although how best to do this is beyond the scope of the present work.

\section{Conclusion}

We have investigated how the \ac{SVD} can be used to reduce the number of
filters needed when analyzing \ac{GW} data for \ac{CBC} signals.  We have found
the number of filters required to matched filter these template banks can be
reduced by about an order of magnitude through truncating the \ac{SVD} of these
waveforms. This result differs from other work that models \ac{CBC} \ac{GW}
signals in approximate ways~\cite{Chassande-Mottin2006, Candes2008,
BuonannoChenVallisneri:2003a} by starting with an exact representation of the
desired template family and producing a rigorous approximation with a tunable
accuracy.

We plan to explore several topics in future works.  Among these are the
derivation of a composite detection statistic using only the \ac{SVD}
coefficients in order to minimize the computational costs associated with
reconstruction and the interpolation of signals not in the original template
set.

\begin{acknowledgments}

The authors would like to acknowledge the support of the LIGO Lab, NSF grants
PHY-0653653 and PHY-0601459, and the David and Barbara Groce Fund at Caltech.
LIGO was constructed by the California Institute of Technology and
Massachusetts Institute of Technology with funding from the National Science
Foundation and operates under cooperative agreement PHY-0757058. The authors
also thank Stephen Privitera and Ik Siong Heng for useful comments and
discussions on this manuscript. This paper has LIGO Document Number
LIGO-P1000037-v2.

\end{acknowledgments}

\bibliography{references}

\end{document}

%% file: packages.tex
\usepackage{acronym}
\usepackage{amsmath}
\usepackage{amssymb}
\usepackage{graphicx}
\graphicspath{{figures/}}
\usepackage{subfigure}
\usepackage[usenames]{color}
\usepackage[normalem]{ulem}
\usepackage[bookmarksnumbered, bookmarksopen, breaklinks, colorlinks]{hyperref} 

%% file: acros.tex
\acrodef{BNS}{binary neutron star}
\acrodef{CBC}{compact binary coalescence}
\acrodef{GW}{gravitational-wave}
\acrodef{PN}{post-Newtonian}
\acrodef{SNR}{signal-to-noise ratio}
\acrodef{SPA}{stationary phase approximation}
\acrodef{SVD}{singular value decomposition}

%% file: inspiral_svd.bbl
\begin{thebibliography}{14}
\expandafter\ifx\csname natexlab\endcsname\relax\def\natexlab#1{#1}\fi
\expandafter\ifx\csname bibnamefont\endcsname\relax
  \def\bibnamefont#1{#1}\fi
\expandafter\ifx\csname bibfnamefont\endcsname\relax
  \def\bibfnamefont#1{#1}\fi
\expandafter\ifx\csname citenamefont\endcsname\relax
  \def\citenamefont#1{#1}\fi
\expandafter\ifx\csname url\endcsname\relax
  \def\url#1{\texttt{#1}}\fi
\expandafter\ifx\csname urlprefix\endcsname\relax\def\urlprefix{URL }\fi
\providecommand{\bibinfo}[2]{#2}
\providecommand{\eprint}[2][]{\url{#2}}

\bibitem[{\citenamefont{Owen}(1996)}]{Owen:1995tm}
\bibinfo{author}{\bibfnamefont{B.~J.} \bibnamefont{Owen}},
  \bibinfo{journal}{Phys. Rev. D} \textbf{\bibinfo{volume}{53}},
  \bibinfo{pages}{6749} (\bibinfo{year}{1996}).

\bibitem[{\citenamefont{Owen and Sathyaprakash}(1999)}]{Owen:1998dk}
\bibinfo{author}{\bibfnamefont{B.~J.} \bibnamefont{Owen}} \bibnamefont{and}
  \bibinfo{author}{\bibfnamefont{B.~S.} \bibnamefont{Sathyaprakash}},
  \bibinfo{journal}{Phys. Rev. D} \textbf{\bibinfo{volume}{60}},
  \bibinfo{pages}{022002} (\bibinfo{year}{1999}).

\bibitem[{\citenamefont{Brady and Ray-Majumder}(2004)}]{bradyraymajumder2004}
\bibinfo{author}{\bibfnamefont{P.~R.} \bibnamefont{Brady}} \bibnamefont{and}
  \bibinfo{author}{\bibfnamefont{S.}~\bibnamefont{Ray-Majumder}},
  \bibinfo{journal}{Classical and Quantum Gravity}
  \textbf{\bibinfo{volume}{21}}, \bibinfo{pages}{S1839} (\bibinfo{year}{2004}),
  \urlprefix\url{arXiv:gr-qc/0405036}.

\bibitem[{\citenamefont{Heng}(2009)}]{heng2008}
\bibinfo{author}{\bibfnamefont{I.~S.} \bibnamefont{Heng}},
  \bibinfo{journal}{Classical and Quantum Gravity}
  \textbf{\bibinfo{volume}{26}}, \bibinfo{pages}{105005}
  (\bibinfo{year}{2009}),
  \urlprefix\url{http://stacks.iop.org/0264-9381/26/i=10/a=105005}.

\bibitem[{\citenamefont{Wen}(2008)}]{wen2008}
\bibinfo{author}{\bibfnamefont{L.}~\bibnamefont{Wen}}, \bibinfo{journal}{Int.
  J. Mod. Phys. D} \textbf{\bibinfo{volume}{17}}, \bibinfo{pages}{1095}
  (\bibinfo{year}{2008}), \urlprefix\url{http://arxiv.org/abs/gr-qc/0702096v2}.

\bibitem[{\citenamefont{Croce et~al.}(2000)\citenamefont{Croce, Demma, Pierro,
  Pinto, and Postiglione}}]{Croce2000}
\bibinfo{author}{\bibfnamefont{R.~P.} \bibnamefont{Croce}},
  \bibinfo{author}{\bibfnamefont{T.}~\bibnamefont{Demma}},
  \bibinfo{author}{\bibfnamefont{V.}~\bibnamefont{Pierro}},
  \bibinfo{author}{\bibfnamefont{I.~M.} \bibnamefont{Pinto}}, \bibnamefont{and}
  \bibinfo{author}{\bibfnamefont{F.}~\bibnamefont{Postiglione}},
  \bibinfo{journal}{Phys. Rev. D} \textbf{\bibinfo{volume}{62}},
  \bibinfo{pages}{124020} (\bibinfo{year}{2000}),
  \urlprefix\url{http://link.aps.org/doi/10.1103/PhysRevD.62.124020}.

\bibitem[{\citenamefont{Mitra et~al.}(2005)\citenamefont{Mitra, Dhurandhar, and
  Finn}}]{finn2005}
\bibinfo{author}{\bibfnamefont{A.~S.} \bibnamefont{Mitra}},
  \bibinfo{author}{\bibfnamefont{S.~V.} \bibnamefont{Dhurandhar}},
  \bibnamefont{and} \bibinfo{author}{\bibfnamefont{L.~S.} \bibnamefont{Finn}},
  \bibinfo{journal}{Phys. Rev. D} \textbf{\bibinfo{volume}{72}},
  \bibinfo{pages}{102001} (\bibinfo{year}{2005}).

\bibitem[{\citenamefont{Allen et~al.}(2005)\citenamefont{Allen, Anderson,
  Brady, Brown, and Creighton}}]{findchirppaper}
\bibinfo{author}{\bibfnamefont{B.~A.} \bibnamefont{Allen}},
  \bibinfo{author}{\bibfnamefont{W.~G.} \bibnamefont{Anderson}},
  \bibinfo{author}{\bibfnamefont{P.~R.} \bibnamefont{Brady}},
  \bibinfo{author}{\bibfnamefont{D.~A.} \bibnamefont{Brown}}, \bibnamefont{and}
  \bibinfo{author}{\bibfnamefont{J.~D.~E.} \bibnamefont{Creighton}}
  (\bibinfo{year}{2005}), \eprint{gr-qc/0509116}.

\bibitem[{\citenamefont{Wainstein and Zubakov}(1962)}]{wainstein:1962}
\bibinfo{author}{\bibfnamefont{L.~A.} \bibnamefont{Wainstein}}
  \bibnamefont{and} \bibinfo{author}{\bibfnamefont{V.~D.}
  \bibnamefont{Zubakov}}, \emph{\bibinfo{title}{Extraction of signals from
  noise}} (\bibinfo{publisher}{Prentice-Hall}, \bibinfo{address}{Englewood
  Cliffs, NJ}, \bibinfo{year}{1962}).

\bibitem[{\citenamefont{Galassi et~al.}(2009)\citenamefont{Galassi, Davies,
  Theiler, Gough, Jungman, Alken, Booth, and Rossi}}]{GSL}
\bibinfo{author}{\bibfnamefont{M.}~\bibnamefont{Galassi}},
  \bibinfo{author}{\bibfnamefont{J.}~\bibnamefont{Davies}},
  \bibinfo{author}{\bibfnamefont{J.}~\bibnamefont{Theiler}},
  \bibinfo{author}{\bibfnamefont{B.}~\bibnamefont{Gough}},
  \bibinfo{author}{\bibfnamefont{G.}~\bibnamefont{Jungman}},
  \bibinfo{author}{\bibfnamefont{P.}~\bibnamefont{Alken}},
  \bibinfo{author}{\bibfnamefont{M.}~\bibnamefont{Booth}}, \bibnamefont{and}
  \bibinfo{author}{\bibfnamefont{F.}~\bibnamefont{Rossi}},
  \emph{\bibinfo{title}{GNU Scientific Library Reference Manual}}
  (\bibinfo{publisher}{Network Theory Ltd}, \bibinfo{address}{United Kingdom},
  \bibinfo{year}{2009}), \bibinfo{edition}{3rd} ed., \bibinfo{note}{for version
  1.12}.

\bibitem[{\citenamefont{LSC}()}]{LAL}
\bibinfo{author}{\bibnamefont{LSC}},
  \urlprefix\url{https://www.lsc-group.phys.uwm.edu/daswg/projects/lal.html}.

\bibitem[{\citenamefont{Chassande-Mottin and Pai}(2006)}]{Chassande-Mottin2006}
\bibinfo{author}{\bibfnamefont{E.}~\bibnamefont{Chassande-Mottin}}
  \bibnamefont{and} \bibinfo{author}{\bibfnamefont{A.}~\bibnamefont{Pai}},
  \bibinfo{journal}{Phys. Rev. D} \textbf{\bibinfo{volume}{73}},
  \bibinfo{pages}{042003} (\bibinfo{year}{2006}),
  \urlprefix\url{http://arxiv.org/abs/gr-qc/0512137}.

\bibitem[{\citenamefont{Candès et~al.}(2008)\citenamefont{Candès, Charlton,
  and Helgason}}]{Candes2008}
\bibinfo{author}{\bibfnamefont{E.~J.} \bibnamefont{Candès}},
  \bibinfo{author}{\bibfnamefont{P.~R.} \bibnamefont{Charlton}},
  \bibnamefont{and} \bibinfo{author}{\bibfnamefont{H.}~\bibnamefont{Helgason}},
  \bibinfo{journal}{Class. Quant. Grav} \textbf{\bibinfo{volume}{25}},
  \bibinfo{pages}{184020} (\bibinfo{year}{2008}),
  \urlprefix\url{http://arxiv.org/abs/0806.4417}.

\bibitem[{\citenamefont{Buonanno et~al.}(2003)\citenamefont{Buonanno, Chen, and
  Vallisneri}}]{BuonannoChenVallisneri:2003a}
\bibinfo{author}{\bibfnamefont{A.}~\bibnamefont{Buonanno}},
  \bibinfo{author}{\bibfnamefont{Y.}~\bibnamefont{Chen}}, \bibnamefont{and}
  \bibinfo{author}{\bibfnamefont{M.}~\bibnamefont{Vallisneri}},
  \bibinfo{journal}{Phys. Rev. D} \textbf{\bibinfo{volume}{67}},
  \bibinfo{pages}{024016} (\bibinfo{year}{2003}), \bibinfo{note}{erratum-ibid.
  74 (2006) 029903(E)}.

\end{thebibliography}
